\documentclass[11pt,a4paper]{article} 
\pdfoutput=1
\usepackage{style}
\usepackage{amsmath}
\usepackage{amssymb}
\usepackage{amsthm}
\usepackage{psfrag}
\usepackage{graphicx}
\usepackage{hyperref}
\usepackage{feynmp}
\usepackage{color}
\usepackage{bm}
\usepackage{multirow}

\newcommand{\be}{\begin{equation}}
\newcommand{\ee}{\end{equation}}
\newcommand{\beqa}{\begin{eqnarray}}
\newcommand{\eeqa}{\end{eqnarray}}

\newcommand\m{\mu}

\newcommand\g{\gamma}
\newcommand\D{\Delta}

\renewcommand\r{\rho}

\newcommand{\HH}{{\cal H}}

\newcommand{\R}{\mathcal{R}}

\def\d{\partial}
\newcommand{\bseq}{\begin{subequations}}
\newcommand{\eseq}{\end{subequations}}

\renewcommand{\ln}{\mathop{\rm ln}\nolimits}

\relax

\title{Cosmology with a light ghost}

\author[a,b,c]{Mikhail M. Ivanov\footnote{\texttt{mikhail.ivanov@cern.ch}}}
\author[a,b]{Anna A. Tokareva\footnote{\texttt{anna.tokareva@epfl.ch}}}

\affiliation[a]{Institute of Physics, LPPC, \'Ecole Polytechnique F\'ed\'erale de Lausanne, \\ \normalsize\it CH-1015, Lausanne, Switzerland}
\affiliation[b]{Institute for Nuclear Research of the
Russian Academy of Sciences, \\ 
\normalsize \it  60th October Anniversary Prospect, 7a, 117312
Moscow, Russia}
\affiliation[c]{Department of Particle Physics and Cosmology, Physics Faculty, Moscow State University, \\
Vorobjevy Gory, 119991, Moscow, Russia}

\abstract{
We study the creation and evolution of cosmological perturbations
in renormalizable quadratic gravity with a Weyl term. 
We adopt a prescription that implies the stability of the vacuum
at the price of introducing a massive spin-two ghost state, 
leading to the loss of unitarity.
The theory may still be predictive regardless the interpretation 
of non-unitary processes provided that their rate is negligible compared to  
the Universe expansion rate.
This implies that the ghost is effectively stable.
In such a setup, there are two scalar degrees of freedom excited during inflation. 
The first one is the usual curvature perturbation whose power spectrum appears to coincide with that of single-field inflation. 
The second one is a scalar component of the ghost encoded in the shift vector of the metric
in the uniform inflaton gauge.
The amplitudes of primordial tensor and vector perturbations are
strongly suppressed.
After inflation the ghost field starts to oscillate and
its energy density shortly becomes dominant in the Universe. 
For all ghost masses allowed by laboratory constraints
ghosts should have ``overclosed" the Universe
at temperatures higher than that of primordial nucleosynthesis.
Thus, the model with the light Weyl ghost is ruled out.
}

\begin{document}

\begin{flushright}
INR-TH-2016-037
\end{flushright}

\maketitle

\section{Introduction}
\label{sec:intro}

General Relativity (GR) is a very 
powerful effective field theory whose predictions were verified at
various scales. However, elegant as it is, GR
cannot be a fundamental theory of gravity since it is not able to account for physics above the Planck energy.
The quest for a consistent theory of quantum gravity is not yet completed, although there exist promising approaches to this problem.
String theory (e.g. \cite{string}) provides important insights 
at the price of great complexity, 
which makes challenging the investigation of observational consequences.

An interesting proposal is Lorentz-violating Ho{\v{r}}ava-Lifshitz gravity \cite{Horava:2009uw}, a branch of which (the so-called projectable model)
provides the only known example of a consistent quantum field theory for the spin-2 
graviton in four dimensions \cite{Barvinsky:2015kil}.  
Unfortunately, this branch does not reproduce GR at low energies within weak coupling
\cite{Blas:2009yd}.
Another, the so-called non-projectable branch of Ho{\v{r}}ava-Lifshitz gravity 
\cite{Blas:2009qj} is not yet rigorously proven to be renormalizable
but is consistent with all current observations 
\cite{Audren:2014hza,Yagi:2013ava}
and has curious implications for 
inflation \cite{Ivanov:2014yla}, dark matter \cite{Blas:2012vn}, and dark energy \cite{Blas:2011en},
see \cite{Blas:2014aca} for a review. 

Another model of quantum gravity has been known for decades.
This is the gravity theory with quadratic curvature invariants developed by Stelle in Refs \cite{Stelle:1977ry,Stelle:1976gc}. 
For some regions of parameter space quadratic gravity even becomes 
asymptotically free and thus
UV complete \cite{Fradkin:1981iu,Fradkin:1981hx,Avramidi:1985ki}. 
Because of the presence of negative-norm states in the spectrum this theory was not very popular,
but regained interest recently \cite{Salvio:2014soa,Einhorn:2014gfa,Donoghue:2016xnh}.

Quadratic gravity is a higher-derivative theory, which is known to be plagued by the 
Ostrogradsky instability at the classical level \cite{Ostrogradsky,Woodard:2015zca}.
At the quantum level the theory possesses a spin-2 state with a negative kinetic term.
The usual quantization prescription implies that the energy of this state is unbounded from below,
which triggers an inevitable vacuum decay.
However, the spin-2 state can be quantized in a way
that avoids this instability by introducing a negative norm \cite{ghosts};
the resulting particle is called ``ghost"  and has a positively-defined energy.
The presence of the ghost
does not allow the theory to be interpreted 
along the lines of unitary quantum mechanics, however,
a consistent interpretation might still exist, see e.g. Refs. 
\cite{Salvio:2015gsi} and references therein.
First, ghosts may be removed from the partition function by some non-trivial boundary conditions
\cite{Maldacena:2011mk}. 
Second, negative-norm states may be not so dangerous 
if for all measurable outcomes of an experiment overall probabilities remain positive and not exceeding unity \cite{Feynman:1984ie}. 
On the other hand, one can think of quadratic gravity as an ``intermediate" effective theory
with a hope that it admits a unitary UV - completion.

In this study we adopt an agnostic point of view about this problem:
we do not know the interpretation of non-unitary processes, 
but quadratic gravity still can be predictive regardless any interpretation provided 
that these processes are negligible during the lifetime of the Universe.
In this case the violation of unitarity, even if observable in principle, 
is suppressed, and thus the theory becomes viable, at least, at the phenomenological level.

As shown below, this setup requires the ghost mass to be smaller than 10 MeV.
This implies that the ghost has non-negligible fluctuations at inflation, which might have 
observational consequences.

Inflationary perturbations in quadratic gravity
in the case of the ghost mass bigger than the Hubble rate were studied in Ref.~\cite{Deruelle:2010kf}. 
Tensor modes were discussed in detail in Ref. \cite{Clunan:2009er} 
for a generic ghost mass. 
This analysis showed that the amplitude of tensor modes is negligible in 
the limit of a vanishing ghost mass.
Inflation in Weyl gravity has been analyzed in \cite{Myung:2014jha,Myung:2015vya},
where it was also shown that amplitudes of tensor and vector perturbations 
are largely suppressed in the limit of a small ghost mass. 
In Ref.~\cite{Myung:2014jha} it was also found that mode functions of the Newtonian gravitational potential are unstable at superhorizon scales, but no interpretation of this fact was given.

In this paper we investigate the dynamics of cosmological 
perturbations in renormalizable quadratic gravity 
both at inflation and late times.
We show that in the conformal Newtonian gauge the ghost field 
acquires a tachyonic mass on the de Sitter background, 
which leads to a superhorizon growth of ghost fluctuations during inflation.
The back-reaction of metric perturbations is not negligible
and it induces a superhorizon growth of inflaton perturbations.
We argue that this instability is a gauge artifact since there is at least one gauge
(namely, the uniform inflaton gauge) free of superhorizon growth. 

We find two scalar modes that get excited and then conserve outside the horizon during inflation.
The first one is the scalar curvature perturbation, whose amplitude 
turns out to coincide with the standard single-field expression.
The second one is a longitudinal part of the shift vector, which 
is related to the ghost degree of freedom. 
At late times, once the Hubble rate gets equal to the ghost mass, this mode 
starts to oscillate and the corresponding energy density shortly becomes dominant in the Universe.
The situation is somewhat close to the cosmological 
``light moduli problem" \cite{Coughlan:1983ci}, 
however, in our case there are crucial differences that make the problem
more severe.
Our main conclusion is that cosmology with a light ghost is incompatible with a healthy 
late-time evolution.

The paper is organized as follows. 
In Section \ref{sec:setup} we discuss quadratic gravity
and the constraints on the ghost mass.
In Section \ref{sec:infl} we explore the generation of scalar perturbations at inflation and 
observational signatures of the model.
In the Section \ref{sec:late} we study the late-time evolution and elaborate on the 
ghost domination.
We discuss our results and conclude in Section \ref{sec:concl}.
In Appendix \ref{app:gauge} we introduce the relations between
perturbations in the conformal Newtonian and the uniform inflaton gauges.
In Appendix \ref{app:comoving} we cross-check the results of the main text in the uniform 
inflaton gauge. 
Finally, in Appendix \ref{app:late} we study qualitatively the equations for cosmological perturbations
at late times, and derive the ghost energy density.

\section{Preliminaries}
\label{sec:setup}

We start with the action of quadratic renormalizable gravity,
\begin{equation}
\label{eq:actR^2}
S= \frac{M_P^2}{2} \int\!d^4x\,\sqrt{-g}\,\left( R
   + \frac{1}{6 \mu^2} R^2
- \frac{1}{2 m^2}
    \,C_{\mu\nu\rho\sigma}\,C^{\mu\nu\rho\sigma}\,   
   \right)\,,
\end{equation}
where $R$ is the scalar curvature, $C_{\mu\nu\rho\sigma}$ is the Weyl tensor and $M_P$ is the reduced Planck mass.\footnote{\label{fn:def}
Units: $ M_P= 1/\sqrt{8 \pi\,G}\simeq 2.4\times 10^{18}$ GeV, $ c = 1 $.
Conventions: $ (-+++) $;
$ R^\mu{}_{\nu\rho\sigma} = \partial_\rho \Gamma^\mu_{\nu\sigma} - \partial_\sigma \Gamma^\mu_{\nu\rho} + \cdots $;
$ R_{\nu\sigma} = R^\mu{}_{\nu\mu\sigma} $;
$ R = g^{\mu\nu} R_{\mu\nu} $;
$ G_{\mu\nu} = R_{\mu\nu} - \frac{1}{2} g_{\mu\nu} R $.
$ C_{\mu\nu\rho\sigma} = R_{\mu\nu\rho\sigma} - \frac{1}{2} (g_{\mu\rho} G_{\nu\sigma} - g_{\mu\sigma} G_{\nu\rho} - g_{\nu\rho} G_{\mu\sigma} + g_{\nu\sigma} G_{\mu\rho}) - \frac{R}{3} (g_{\mu\rho} g_{\nu\sigma} - g_{\mu\sigma} g_{\nu\rho}) $.
Greek indices run from $ 0 $ to $ 3 $; latin indices run from $ 1 $ to $ 3 $.
} 
Upon flat space this theory 
possesses a massless transverse-traceless spin-2 excitation (graviton), 
a spin-0 particle of mass $\mu$ (dubbed scalaron \cite{Starobinsky:1980te})
and a spin-2 particle of mass $m$ with the ``wrong" sign of its quadratic lagrangian.
In what follows, we adopt a negative-norm prescription for this state, which 
implies the loss of unitarity but ensures the stability with respect to the 
vacuum decay.
In this paper we assume that the ghost and scalaron masses are not tachyonic,
\be 
m^2>0 \quad \text{and} \quad  \mu^2>0\,,
\ee
i.e. both degrees of freedom have standard dispersion relations upon flat space,\footnote{This means that the mass term of the ghost should have the sign opposite to that of a healthy bosonic field.}
\be
\begin{split}
\text{scalaron:}&\quad \omega^2=k^2+\mu^2\,,\\
\text{ghost:}&\quad \omega^2=k^2+m^2\,.
\end{split} 
\ee

Although the condition $\mu^2>0$ corresponds to a region of parameter space
incompatible with asymptotic freedom \cite{Salvio:2014soa,Einhorn:2014gfa},
the case of $\mu^2<0$ is not viable as the cosmological evolution
leads either to a collapse or to a run-away
singularity \cite{Ruzmaikin,Barrow:2006xb,Ivanov:2011np}.

It is convenient to switch to the Einstein frame by means of a conformal transformation \cite{Magnano:1987zz}. 
Recall that the square of the Weyl tensor is an invariant under the conformal symmetry. This yields the following action:
\be
\begin{split}
S&=S_{GR}+S_{\phi}+S_{\mathrm{Weyl}}\\
&= \frac{M_P^2}{2 } \int\!d^4x\,\sqrt{-g}\,R
  - \frac{1}{2}
    \int\!d^4x\,\sqrt{-g}\,
    \left(\partial_\mu \phi\,\partial^\mu \phi + 2 V(\phi)\right)
  - \frac{M_P^2}{4 m^2}
    \int\!d^4x\,\sqrt{-g}\,C_{\mu\nu\rho\sigma}\,C^{\mu\nu\rho\sigma}\,,
\label{eq:action}
\end{split}
\ee
where $V(\phi)$ is the Starobinsky potential,
\be
V(\phi)= \frac{3}{4}\mu^2 M_P^2\left(1-e^{-\sqrt{\frac{2}{3}}\frac{\phi}{M_P}}\right)^2\,.
\ee
Although the Starobinsky model \cite{Starobinsky:1980te}\footnote{For the dynamics in the Einstein frame see Ref.~\cite{DeFelice:2010aj}.} 
 provides us with the most natural way to organize inflation in 
quadratic gravity, in our analysis we will assume a generic potential $V(\phi)$ with the
requirements that it allows for a slow-roll regime and may be obtained from a renormalizable theory.
In this way our results will be more model-independent, and applicable, for instance,
to agravity-based inflationary models discussed in Refs.~\cite{Salvio:2014soa,Kannike:2015apa}.
Note that if we had not required renormalizability, the
model with a generic potential would, formally, be equivalent to $f(R)$-gravity plus a Weyl term.
We will refer to field $\phi$ as ``inflaton" in what follows.

The variation of \eqref{eq:action} with respect to the metric and $\phi$ yields the system of the modified Einstein and Klein-Gordon-Fock equations,
\be 
\label{eq:eqgen}
\begin{split}
& \nabla^{\mu}\nabla_{\mu}\phi -V'(\phi)=0\,,\\
&R_{\mu \nu}-\frac{1}{2}Rg_{\mu\nu}-\frac{1}{m^2}B_{\mu \nu}=\frac{1}{M_P^2}T_{\mu \nu}\,,\\
&\quad \text{where}\quad  B_{\mu \nu}\equiv   \left(2\nabla^{\alpha} \nabla^{\beta} +
R^{\alpha\beta}-\frac{1}{2}g^{\alpha \beta}R\right)C_{\mu \alpha \nu \beta}\,,\\
& \quad \quad \quad  \quad \quad T_{\mu \nu}=\d_\mu \phi \d_\nu \phi -g_{\mu \nu}\left(\frac{1}{2}(\d_\mu \phi)^2+V(\phi)\right)\,,
\end{split}
\ee
and $\nabla_{\mu}$ is the covariant derivative.

The presence of negative-norm states implies processes with negative widths and cross-sections. 
Their interpretation is still an open issue, e.g. 
a negative decay width might actually signalize the violation of causality \cite{Grinstein:2008bg}.
As shown in this reference, even unitarity may be restored within
the acausal interpretation. 
It is not clear, however, how to extend this interpretation 
to account for non-unitary processes in cosmology.
On the other hand, one can try to adopt the ``usual" probabilistic interpretation, 
which suggests that negative-probability processes are instabilities.
Indeed, in such a setup the ``ghost decay" implies
an exponential growth of ghost particles' energy density, which follows from the Boltzmann equation,
\be 
\frac{d n_{gh}}{dt}+3H n_{gh}=|\Gamma | n_{gh}\,,
\ee
 where $H$ is the Hubble rate, $n_{gh}$ is the ghost number density, and $\Gamma$ is the (negative) decay rate of ghosts.\footnote{It should be pointed out that there can be many other ways 
to interpret theories with negative-norm states~\cite{Salvio:2015gsi}. 
For instance, one can define probabilities to be always positive,
but then their sum will exceed unity. In this case negative states do not imply negative probabilities.}

The scalaron and the ghost are identified with metric components and thus
couple to matter universally (see e.g. Ref.~\cite{Gorbunov:2010bn}). 
The expression for the decay width of ghosts is similar (up to a numerical factor) 
to that of the scalaron \cite{starobinsky,Vilenkin:1985md}, 
but with a minus sign originating from the negative-norm prescription,
\begin{equation}
\Gamma \sim -\frac{m^3}{M_P^2}\,.
\end{equation}
One can suppress this process
by properly tuning the ghost mass. 
Requiring the characteristic time of ghost decay to be larger than the 
lifetime of the Universe one gets a conservative bound 
\be 
 m \lesssim 10 \; \text{MeV}. 
\ee

The lower bound on the ghost mass comes from short-scale gravity experiments. 
In the Newtonian limit the action \eqref{eq:actR^2} yields the usual GR gravitational 
potential plus corrections 
suppressed by Yukawa-type exponents \cite{Stelle:1977ry}. 
The bounds on the latter \cite{Geraci:2008hb,Tan:2016vwu}
imply the following double-sided constraint on 
the ghost mass:
\be 
\label{eq:range}
10^{-1}\;\mathrm {eV} \lesssim m \lesssim 10 \; \mathrm{MeV}.
\ee

To summarize, let us outline different regimes of quadratic gravity 
depending on the ghost mass $m$ and its relation to the Hubble rate at inflation $H_{inf}$,
\begin{itemize}
\item $m>H_{inf}$ (``heavy ghost"). The ghost is gapped at inflation and has no effect on cosmology. 
Note that adimensional gravity requires the ghost mass to be smaller than $3\times 10^{10}$ GeV
for the naturalness of the small electroweak scale \cite{Salvio:2014soa}.
Since for simplest inflationary scenarios $H_{inf}\gtrsim 10^{13}$ GeV,
achieving the naturalness might be challenging in agravity with a ``heavy ghost".\footnote{
Among all of scenarios studied in Ref.~\cite{Kannike:2015apa},
the lowest possible Hubble scale at inflation  
is given by that of Starobinsky inflation, $H\sim 10^{13}$ GeV.}

\item $10$ MeV $<m\ll H_{inf}$ (``intermediate mass ghost").
The ghost is gapless at inflation and thus its fluctuations get excited. 
For this mass range non-unitary processes (like the ghost decay) happen after inflation 
and thus the viability of the theory depends on their interpretation.
Since a consistent interpretation does not exist at the moment, 
phenomenological implications are speculative. 
Recall that if $m\lesssim 10^{10}$ GeV, the small electroweak scale is natural in agravity. 
Note that our analysis of inflation (Sec.~\ref{sec:infl}) is applicable to this mass range.

\item $0.1$ eV $\lesssim m \lesssim 10$ MeV (``light ghost").  
The ghost decay is suppressed and the laboratory constraints are satisfied.
Quadratic gravity may be
phenomenologically viable and its predictions are not sensitive to any
interpretation of non-unitary processes. We will focus on this regime and study the theory both at
inflation (Sec.~\ref{sec:infl}) and late times (Sec.~\ref{sec:late}).
\end{itemize}

\section{Inflation}
\label{sec:infl}

We start with the creation of cosmological perturbations at inflation. 
Since the Weyl term disappears on the homogeneous and isotropic metric,
the background inflationary dynamics in our model is similar
to that of single-field inflation.
By plugging the homogeneous and isotropic ansatz,
\be
\begin{split}
&ds^2=a(\eta)^2(-d\eta^2+{\bf dx}^2)\,,\\
&\phi=\phi(\eta)\,,
\end{split} 
\ee
into Eq.~\eqref{eq:eqgen}, we obtain the usual background equations:
\be
\label{eq:backgr}
\begin{split}
&3M_P^2\HH^2=\frac{\phi'^2}{2}+a^2(\eta)V\,,\\
&3M_P^2\HH'=-\phi'^2+a^2(\eta)V\,,\\
&\phi''+ 2\HH\phi'+a^2V_{,\phi}=0\,,
\end{split} 
\ee
where primes denote partial derivatives w.r.t. conformal time, 
$'\equiv \partial/ \d \eta$, and $\HH\equiv a'/a$ is the conformal Hubble parameter.
Where convenient, we will also use the quantities 
expressed in the cosmic time $t=\int a(\eta)d\eta$,
and denote the derivatives w.r.t. this time with dots, e.g.
$H\equiv {\dot a}/{a}$.

The accelerated expansion occurs as long as the slow-roll parameters are small. 
The latter are defined in a usual way,
\be
\begin{split}
\label{eq:eps}
&\epsilon_1\equiv\frac{M_P^2}{2}\left(\frac{V_{,\phi}}{V}\right)^2=\frac{\phi'^2}{2\HH^2 M_P^2}
\,,\\
&\epsilon_2\equiv \frac{M_P^2 V_{,\phi\phi}}{V}
\,.
\end{split}
\ee
In the formal limit $V\to$ const (equivalently, $\phi'\to 0$) slow-roll parameters vanish and 
the background reduces to exact de Sitter space characterized by the scale factor
\be 
a(\eta)=-1/(H\eta)\,,
\ee
with $H= M_P^{-1}\sqrt{V/3}=$ const.

Now we focus on linear scalar perturbations in the conformal Newtonian gauge, 
\be 
\begin{split}
\label{metric}
&ds^2
= a(\eta)^2\,
  \left[
   -(1 + 2 \Psi(\eta,{\bf x}))\,d\eta^2 
   + (1 + 2 \Phi(\eta,{\bf x}))\,\delta_{ij} \,dx^i\,dx^j \right]
 \,,\\
&
\phi=\phi(\eta)+\varphi(\eta,{\bf x})\,.
\end{split}
\ee
Expanding the action 
\eqref{eq:action} 
to quadratic order one finds,\footnote{Here we omit the terms that vanish due to the background equations \eqref{eq:backgr}.}
\begin{subequations}
\label{eq:actpert}
\begin{align}
\label{eq:weyl2}
&S_{\text{Weyl}}^{(2)}=
  - \frac{M_P^2}{3 m^2}
    \int\!d^4x\,
    [\Delta (\Psi-\Phi)]^2\,,\\
    \label{eq:gr2}
&S^{(2)}_{GR}
= \frac{M_P^2}{2} \int\!d^4x\,a^2\,
  \left[-6 \Phi'^2
   + 12 \mathcal H\,\Psi\,\Phi'
   - 4 \Phi\Delta \Psi
     -2 \Phi\Delta \Phi 
   - 2     (\mathcal H' + 2 \mathcal H^2)\,\Psi^2\right]\,, \\
&
\label{eq:infl2}
S_{\phi}^{(2)}= \frac{1}{2}\int\!d^4x\,a^2\,
    \left(\varphi'^2
     +\varphi \Delta \varphi
     - a^2\,V_{,\phi\phi}\,\varphi^2
     +6 (\phi'\,\varphi'-V'\varphi)\, \Phi
     - 2 (\phi'\,\varphi'\,
     + a^2\,V_{,\phi}\,\varphi) \Psi \right)\,,
\end{align}
\end{subequations}
where $\Delta \equiv \d^i\d_i$ stands for a spatial Lapacian.

In our analysis we adopt the following strategy. 
We first consider the limit of exact de Sitter 
space suggested by the smallness of the slow-roll parameters. 
In this limit the inflaton decouples from the metric, 
so the inflaton and the ghost behave as free fields in an external background.
At the next step we allow for the finite slow roll-parameters and compute corrections due to the inflaton-metric mixing.

\subsection{Exact de Sitter background}
\label{Sec-2-2}

We start the analysis with the limit of exact de Sitter space, in which 
the inflaton decouples 
from metric perturbations.
The latter get excited too 
since they contain the ghost degree of freedom. 
We will focus on its scalar polarization, which will be referred to as just ``ghost" 
whenever there is no possible confusion.
The quadratic action for the inflaton fluctuations \eqref{eq:infl2} then takes the usual form:
\begin{equation}
\label{eq:actinfds}
S^{(2)}_{\phi}=\frac{1}{2}\int d\eta d^3 x \;a^2 \Big[ (\varphi')^2 - (\partial_i \varphi)^2 \Big]\,.
\end{equation}
One quantizes the inflaton as usual,
\be 
\label{eq:harm1}
\varphi({\bf x},\eta)=\int \frac{d^3k}{(2\pi)^3}(\varphi_{\bf k}(\eta)a_{\bf k}+\varphi^*_{\bf k}(\eta)a^+_{-\bf k})e^{i{\bf kx}}\,,
\ee
with the creation-annihilation operators satisfying 
\be 
[a_{\bf k}, a^+_{\bf q}]=(2\pi)^3\delta^{(3)}({\bf k}-{\bf q})\,.
\ee
The mode functions $\varphi_{\bf k}$ are positive frequency solutions of equation
\be
\label{eq:dseom}
\varphi''_{\bf k}+2\HH  \varphi'_{\bf k}+k^2\varphi_{\bf k}=0\,.
\ee
They have the standard form:
\be
\label{eq:dsinf}
\varphi^{(dS)}_{\bf k}\equiv  \frac{H}{\sqrt{2k}}|\eta|\left(1-\frac{i}{k\eta}\right)e^{-ik\eta}\,.
\ee

\subsubsection*{Ghost perturbations}

\begin{figure}[h]
\begin{center}
\includegraphics[width=0.7\textwidth]{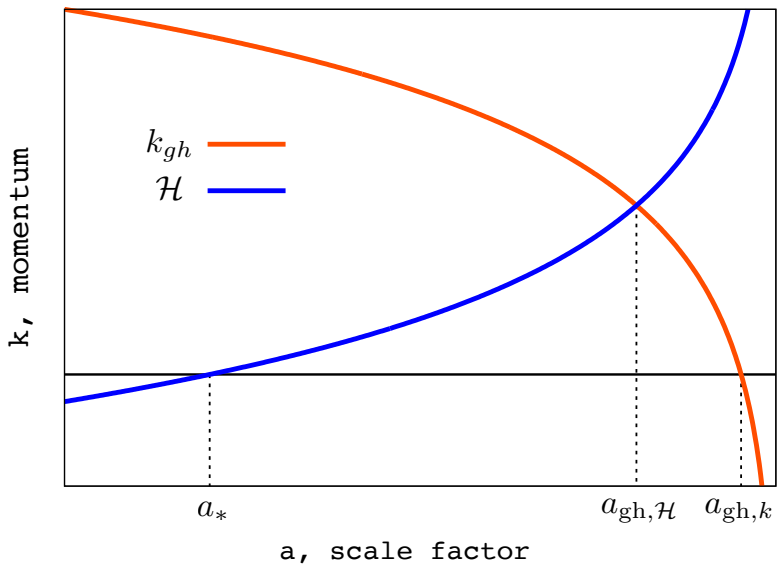}
\end{center}
\caption{\label{fig:kgh}
The time dependence of dynamical scales defining the evolution 
of ghost perturbations: a conformal momentum $k$ (horizontal solid black line),
the conformal Hubble parameter $\HH$ (solid blue line), 
and the ghost scale $k_{gh}=k^2/(ma(\eta))$ (solid orange line).
For other notations see the main text.
}
\end{figure}

Now let us focus on the metric components.
The field $\Psi$ entering Eqs.~\eqref{eq:weyl2} and~\eqref{eq:gr2} is non-dynamical since
it has only one derivative acting on it. 
Varying the actions \eqref{eq:weyl2} and \eqref{eq:gr2} with respect to $\Psi$, and 
performing the Fourier decomposition $\Phi,\Psi \propto e^{i{\bf kx}}$, one finds 
the following constraint equation:
\be 
\label{eq:constr}
-\frac{2}{3m^2a^2}k^4(\Psi-\Phi)+6\HH \Phi'+2k^2 \Phi-2(2\HH^2+\HH')\Psi=0\,.
\ee
One observes the appearance of a new time-dependent scale 
$k_{gh}(\eta)\equiv k^2/(a(\eta)\,m)$
which also depends on Fourier harmonics. 
We call this scale {\it ghost horizon} in what follows.
In order to understand the role of this scale, 
consider the dynamics of a Fourier mode with conformal momentum $k$ (see Fig.~\ref{fig:kgh}).
The evolution of this mode starts deep inside the Hubble horizon, when
the ghost scale is parametrically enhanced compared to $k$. 
In other words, the mode $k$ is outside the ghost horizon.
In this regime the leftmost term in Eq.~\eqref{eq:constr} dominates. 
Since this terms originates from the Weyl tensor squared in the action,
one may say that the dynamics is governed by the Weyl term.
At the Hubble crossing, $\HH=k$, the mode $k$ is still well outside the 
ghost horizon. As the Universe expands, the ghost scale decreases
and eventually gets smaller than the Hubble rate; after that the dynamics of metric perturbations 
is dominated by terms that originate from the Einstein-Hilbert action.\footnote{Recall that we are working in the Einstein frame, in which the theory with $R^2$ in the action
is equivalent to Einstein gravity plus a minimally coupled scalar field.}
The ghost horizon gets equal to the Hubble rate 
after $\sim 15$ e-foldings
since the mode has left the Hubble horizon,
\be 
\ln \frac{a_{\text{gh},\HH}}{a_*}=\ln \sqrt{\frac{H}{m}}\simeq
17 
 \times\ln \left(\left[\frac{H}{10^{13}\;\mathrm{GeV}}\right]\left[\frac{10 \;\mathrm{MeV}}{m}\right]\right)\,,
\ee
where $a_{\text{gh},\HH}$ is the scale factor at which the Hubble rate starts to 
dominate over the ghost scale,
and $a_*$ is the scale factor at the Hubble crossing.

Eventually, the ghost scale gets smaller than the momentum $k$ 
and the corresponding mode enters the ghost horizon. This happens after 
$\sim 30$ e-foldings after the Hubble crossing,
\be 
\ln \frac{a_{\text{gh},k}}{a_*}=\ln \frac{H}{m}\simeq
34 
 \times\ln \left(\left[\frac{H}{10^{13}\;\mathrm{GeV}}\right]\left[\frac{10 \;\mathrm{MeV}}{m}\right]\right)\,,
\ee
where $a_{\text{gh},k}$ is the scale factor at the ghost horizon crossing.
Let us discuss the different dynamical regimes separately.

{\bf Weyl term domination,} $k_{gh}\gg \HH,k$.
The solution of Eq. \eqref{eq:constr} at leading order in $\HH/k_{gh},k/k_{gh}$ is given by
\be
\label{eq:phipsi}
\Psi=\Phi 
\,,
\ee
which, upon substituting back into Eqs.~\eqref{eq:weyl2} and \eqref{eq:gr2}, and 
canonical normalization $\hat\Phi\equiv M_P\sqrt{6} \Phi$,
yields the following quadratic action for the $\Phi$ field,
\be
\label{eq:norm}
S_{\hat{\Phi}}^{(2)}\Big|_{k,\HH\ll k_{gh}}= \frac{1}{2}\int d^4x\; a^2\left[-\hat{\Phi}'^2+(\d_i\hat \Phi)^2-\frac{4}{3}(2\HH^2+\HH')\hat\Phi^2\right]\,.
\ee
We observe that the scalar part of the massive spin-2 ghost 
acquires a tachyonic mass on the de Sitter background.
In order to quantize the ghost
we introduce usual Fourier harmonics analogous to \eqref{eq:harm1},
\be 
\label{eq:dec}
\hat \Phi({\bf x},\eta)=\int \frac{d^3k}{(2\pi)^3}(\hat \Phi_{\bf k}(\eta)b_{\bf k}+\hat \Phi^*_{\bf k}(\eta)b^+_{-\bf k})e^{i{\bf kx}}
\ee
with the creation-annihilation operators satisfying\footnote{
We assume that $b_{\bf k}$ is the annihilation operator for 
the Bunch-Davies vacuum,  i.e. $b_{\bf k}|0\rangle=0$.
} 
\be 
\label{eq:negative}
[b_{\bf k}, b^+_{\bf q}]=-(2\pi)^3\delta^{(3)}({\bf k}-{\bf q})\,.
\ee
Notice that these operators commute with those introduced in Eq.~\eqref{eq:harm1},
\be
\label{eq:zerocom} 
[a^+_{\bf k}, b_{\bf q}]=[a_{\bf k}, b^+_{\bf q}]=0\,.
\ee
The mode functions $\hat \Phi_{\bf k}$ satisfy the equation of motion derived from Eq.~\eqref{eq:norm}, 
\be
\label{eq:eomgh}
\hat\Phi''_{\bf k}+2\HH \hat\Phi'_{\bf k}+k^2 \hat\Phi_{\bf k}-4\HH^2 \hat\Phi_{\bf k}=0\,,
\ee
where we made use $\HH'=\HH^2$ for de Sitter space.
The positive-frequency solution of Eq.~\eqref{eq:eomgh} satisfying the Bunch-Davies initial condistions is,
\be
\label{eq:ghds}
\hat \Phi_{\bf k} = \frac{H}{\sqrt{2k}} |\eta|\left(1-\frac{3i}{k\eta}-\frac{3}{(k\eta)^2}\right)e^{-ik\eta}\,.
\ee
One observes that the tachyonic mass of the ghost leads to a superhorizon 
instability developing as the scale factor.
This instability signalizes the breakdown of 
perturbation theory in the Newtonian gauge. 
Moreover, the superhorizon growth of ghost perturbations
should backfire on the inflaton fluctuations due to the mixing.
Nevertheless, let us proceed and see where computations lead to.

{\bf Einstein term domination,} $\HH \gg k_{gh}$.
In this case the Weyl term becomes negligible and 
the solution of Eq.~\eqref{eq:constr} at next-to-leading order in $k_{gh}/\HH$ is given by
\be
\label{eq:constout}
 \Psi=\frac{\Phi'}{\HH}+\frac{k^2\Phi}{3\HH^2} -\frac{k^2_{gh}(\eta)}{9 \HH^2}\left(\frac{\Phi' }{\HH}+\frac{k^2\Phi}{3\HH^2}-\Phi\right)\,.
\ee
Switching to real space and plugging the above solution back in the actions \eqref{eq:weyl2}, 
\eqref{eq:gr2} yields
\be
 S_{\Phi}^{(2)}\Big|_{\HH\gg k_{gh}}=-\int d^3xd\eta\; \frac{M_P^2\Delta^2}{3 m^2}\left[\frac{\Phi'}{\HH}-\frac{\Delta \Phi}{3\HH^2}-\Phi\right]^2 \,.
\ee
The above action implies the following equation of motion in the superhorizon limit,
where one can neglect spatial laplacians,
\begin{equation}
\Phi''+\frac{2}{\eta}\Phi'=0.
\end{equation}
This equation has a growing solution, 
\be 
\label{eq:Pgr}
\Phi\propto \eta^{-1}\sim a(\eta)\,,
\ee
which accords with the leading superhorizon behavior of the mode function \eqref{eq:ghds}.
Thus, mode functions of $\Phi$ do not change as the Hubble rate starts to dominate over
the ghost scale.
Substituting the solution \eqref{eq:Pgr} into Eq.~\eqref{eq:constout} yields, at the leading order in
$k^2/\HH^2$,
\be
\label{eq:psipphi2}
\Phi=\Psi\,. 
\ee
Remarkably, the solutions \eqref{eq:ghds} and \eqref{eq:phipsi} do not change when 
the Weyl terms becomes subdominant. In other words, 
the ghost perturbations do not ``feel" 
that the Einstein term started to dominate.
This is not accidental, the condition \eqref{eq:phipsi} (\eqref{eq:psipphi2})
reflects the fact that scalar perturbations on exact de
Sitter space are conformally flat, in which case the Weyl tensor squared \eqref{eq:weyl2} vanishes.
The existence of this mode was obtained for the first time in Ref.~\cite{Starobinsky:1981zc}. This mode also appears in some ghost-free non-local 
generalizations of quadratic gravity, e.g. \cite{Koshelev:2016xqb}.

To sum up, we have shown that the constraint \eqref{eq:phipsi} and the
solution for ghost mode functions \eqref{eq:ghds}
are valid all the time during inflation, 
even though the terms that dominate in the action change.

\subsection{Beyond the limit of exact de Sitter space}

In this subsection we account for corrections
due to the inflaton-ghost mixing. 
Within our scheme,
the metric perturbations
obtained in the previous subsection act as sources 
in the inhomogeneous equation of motion for the inflaton derived from the action \eqref{eq:infl2},
\be
\label{eq:phis}
\begin{split}
\varphi''_{\bf k}+2\HH \varphi'_{\bf k}+k^2 \varphi_{\bf k} &=-2\phi' \Phi'_{\bf k}-2a^2 V_{,\phi}\Phi_{\bf k} \,.
\end{split} 
\ee
The field $\Phi$ is given by Eq.~\eqref{eq:ghds}. 
Its has the following leading superhorizon asymptotic:
\be
\label{eq:phiasymp}
\Phi_{\bf k}\Big|_{\eta \to -0}=-\frac{3 H^2}{2\sqrt{3\cdot k}k^2 M_P} a(\eta)e^{-ik\eta}\equiv 
C_{\bf k} a(\eta)\,,
\ee
where for convenience we switched back to a dimensionless variable and
factorized the time dependence.
Eq.~\eqref{eq:phis} with the source \eqref{eq:phiasymp} has the following solution:
\be
\label{eq:modeSR}
\varphi_{\bf k}=\varphi^{(dS)}_{\bf k}+\frac{2(\phi''+\HH\phi')}{3\HH^2+\HH'} C_{\bf k}a(\eta)\,.
\ee

Using the conditions $\ddot\phi\approx  0$, $\HH^2\approx \HH'$ valid in the slow-roll approximation,
Eq. \eqref{eq:modeSR} reads
\be
\label{eq:phitot}
\varphi_{\bf k}=\varphi^{(dS)}_{\bf k}+\frac{\phi'}{\HH} C_{\bf k}a(\eta)\,.
\ee
One concludes that both the inflaton and ghost perturbations in the 
Newtonian gauge grow outside the horizon.
This growth may be interpreted as a gauge artifact
because all perturbations remain bounded in the 
uniform inflaton gauge.

\subsubsection*{Uniform inflaton gauge}

There are three scalar modes of the metric in the uniform inflaton (comoving slicing) gauge,
\begin{equation}
\label{eq:comg0}
\quad g_{00}=a^2(1+2A),~ g_{0i}= a^2 \partial_i B,~g_{ij}=a^2\delta_{ij}(1+2 \R)
\,.
\end{equation}

The gauge invariant quantity $\R$
defined as a perturbation of the number of e-foldings
on the surfaces of the 
uniform inflaton field can be retrieved from the modes of the Newtonian gauge via
\be
\R= \Phi-\frac{\HH}{ \phi'}\varphi\,,
\ee
see App. \ref{app:gauge} for more detail.
One can define mode functions of $\R$ similarly to Eq.~\eqref{eq:harm1}
and compute them using the mode functions for $\varphi$ and $\Phi$ (Eqs.~\eqref{eq:eomgh} and~\eqref{eq:phitot}), which yields
\be 
\label{eq:zeta1}
\mathcal{R}_{\bf k}=\Phi_{\bf k}-\frac{\HH}{ \phi'}\varphi_{\bf k}=-\frac{\HH}{\phi'}\varphi^{(dS)}_{\bf k}
=-\frac{H}{\dot \phi}\varphi^{(dS)}_{\bf k}\,,
\ee
where $\varphi^{(dS)}_{\bf k}$ is given by Eq.~\eqref{eq:dsinf}.
One observes that the growing contributions cancel each other in Eq.~\eqref{eq:zeta1},
so the curvature perturbation coincides with that of the single-field case. Thus, it
remains constant outside the horizon analogous to the case of usual gravity \cite{Bardeen:1983qw,Salopek:1990jq}.
In App.~\ref{app:comoving} we additionally verify that $\R$ is constant 
outside the horizon by a direct calculation in the uniform inflaton gauge.
Note that the constancy of $\R$ in quadratic gravity 
was proven for the first time in Ref.~\cite{Starobinsky:2001xq} (see also Ref.~\cite{Deruelle:2010kf}).

Another non-trivial scalar perturbation in the uniform inflaton 
gauge is the longitudinal part $B$ of the shift vector $ g_{0i}$, see Eq.~\eqref{eq:comg0}.
The corresponding mode functions can be retrieved from the inflaton perturbation in the Newtonian
gauge via
\be 
\label{eq:B1}
\begin{split}
& B_{\bf k}=\frac{\varphi_{\bf k}}{\phi'}
=\R_{\bf k}\eta+\frac{C_{\bf k}}{H}
\,,
\end{split}
\ee
where in the last equality we have used Eqs.~\eqref{eq:phitot} and \eqref{eq:zeta1}.
The first contribution in the r.h.s. of Eq.~\eqref{eq:B1}(adiabatic mode) vanishes at $\eta\to 0$. 
The second contribution is induced by the ghost perturbations 
and is constant in the limit $\eta \to 0$,
\be
B_{0,{\bf k}} \equiv \frac{C_{\bf k}}{H}=-\frac{\sqrt{3}}{2 k^{5/2}}\frac{H}{M_P}\,.
\ee
The gauge invariant generalisation of $B$ 
is constructed in App.~\ref{app:gauge}.

Finally, the lapse function $A$ vanishes at superhorizon scales,
\be
\label{eq:lapse}
A=\Psi-\HH B-B' =\R-B'=0\,,
\ee
where we used Eqs.~\eqref{eq:phipsi},~\eqref{eq:zeta1}, and \eqref{eq:B1}.

We have seen that inflationary perturbations are plagued by 
superhorizon growth in the conformal Newtonian gauge. This growth,
however, should be treated as unphysical
since all perturbations are bounded and conserved outside the horizon 
in the comoving slicing gauge.
This suggests two ways to perform calculations: 

1) remain in the Newtonian gauge and ignore the 
fact that growing superhorizon modes break perturbation theory.
These modes should eventually cancel in gauge-invariant quantities.

2) work in the comoving slicing gauge from the beginning.

In this section we followed the first way. 
In App. \ref{app:comoving} we perform the analysis directly in the comoving slicing and 
show that the two approaches yield the same results.

\subsection{Observational predictions}

The power spectrum of curvature perturbations is given by a standard 
expression,
\be
\label{eq:Pr}
\begin{split}
 \mathcal{P}_{\R}(k)
 =
 \frac{k^3}{2\pi^2}\lim_{\eta\to -0}|\R_{\bf k}(\eta)|^2
=\frac{H^4}{4\pi^2\dot\phi^2}\Bigg|_{aH=k}\,,
\end{split}
\ee
where we 
used Eqs.~\eqref{eq:zeta1} and \eqref{eq:dsinf}, and
explicitly emphasized that the spectrum above should be evaluated at the horizon crossing.

As for the conserved scalar perturbation $B$, its power spectrum is given by,
\be 
\label{eq:Bspectr}
\langle |B(\eta,{\bf x})|^2\rangle
\underset{\eta \to -0}\longrightarrow 
\langle |B_0({\bf x})|^2\rangle
=
\int \frac{dk}{k}\frac{\mathcal{P}_B(k)}{k^2}\,,
\ee
where $\mathcal{P}_B(k)$ is the power spectrum of the shift vector (gradient of $B$),
which closely resembles the power spectrum of tensor 
modes in single-field inflation,
\be 
\begin{split}
\mathcal{P}_{B}(k)
 =
 \frac{k^5}{2\pi^2}|B_{0,{\bf k}}|^2
=\frac{3H^2}{8\pi^2 M_P^2}\Bigg|_{aH=k}\,.
 \end{split}
\ee
Notice that the amplitude \eqref{eq:Bspectr} is quadratically divergent for long-wavelength perturbations.
This is not so worrisome because 
there exists a natural infrared cutoff set by the Hubble scale.
Modes larger than the Hubble horizon cannot be 
observed locally and 
can be removed by a suitable gauge transformation, thus they
do not contribute to the local energy density. 
Situation here is similar to the case of primordial
gravitational waves, even though their amplitude is 
only logarithmically divergent for a spectrum close to scale-invariant
\cite{Starobinsky:1979ty,Allen:1987bk,Gorbunov:2011zzc}.

In order to study the correlation between the fields $B$ and $\R$
we express them through the inflaton $\varphi$ and the Newtonian potential $\Phi$ as
in Eqs.~\eqref{eq:zeta1} and~\eqref{eq:B1}.
Making use of the decompositions \eqref{eq:dec} and \eqref{eq:harm1}, and Eq.~\eqref{eq:zerocom},
we obtain that the correlation between two fields vanishes at superhorizon scales,
\be
\langle B(\eta,{\bf x})\R(\eta,{\bf x}) \rangle =
-\frac{\HH}{\phi'^2}\langle |\varphi(\eta,{\bf x})|^2 \rangle
=\eta \int \frac{dk}{k}  \mathcal{P}_\R 
\underset{\eta \to -0}\longrightarrow 0\,.
\ee

The power spectrum of tensor modes has been computed 
in Refs.~\cite{Clunan:2009er,Myung:2014jha,Myung:2015vya},
\be
\label{eq:Ph}
\mathcal{P}_h=\frac{2}{\pi^2}\frac{H^2}{M_P^2}\frac{1}{1+2\frac{H^2}{m^2}}\Bigg|_{aH=k}\simeq \frac{1}{\pi^2}\frac{m^2}{M_P^2}\Bigg|_{aH=k}\,.
\ee
Note that this result is very similar (up to a numerical factor) 
to the expression for the spectrum of gravitational waves 
in Starobinsky inflation \cite{Starobinsky:1983zz}, which is recovered
by changing $m\to \mu$ (the scalaron mass).
Values of the ghost mass lying in the range
\eqref{eq:range} imply an extremely small scalar-to-tensor ratio with no hope for detection,
\be
\label{eq:r}
r=\frac{\mathcal{P}_h}{\mathcal{P}_\R}=7\times  10^{-34}\left[\frac{m}{10 \;\mathrm{MeV}}\right]^2\left[\frac{10^{-10}}{\mathcal{P}_\R}\right]\,.
\ee
The limit $m\to 0$ corresponds to pure conformal gravity
where de Sitter space is equivalent to flat space,
in which the amplification of tensor modes does 
not take place \cite{Starobinsky:1981zc}.
The spectrum of vector modes was found to 
be equal to \eqref{eq:Ph} \cite{Myung:2014jha} and
hence is also vanishing.

To sum up, the model predicts the same scalar curvature perturbations as usual inflation,
the existence of scalar ghost perturbations 
and vanishing tensor and vector modes.
A non-trivial step in analyzing the scalar sector is the choice of the gauge.
We have shown that the conformal Newtonian gauge is unstable because of the 
tachyonic mass of the ghost. 
The fluctuations behave well in the comoving slicing, 
although the lagrangian for perturbations is 
(approximately) diagonal and 
can be conventionally quantized only in the Newtonian gauge
(see App.~\ref{app:comoving} for more details).

\section{After inflation: the cosmological ghost problem}
\label{sec:late} 

Reheating after inflation takes place during the oscillations of the inflaton field, 
which cause the production of light particles and can be seen 
as the inflaton decay.
In our setup the ghost is too light to decay 
and thus reheating in our model is completely identical that of single-field inflation.
From now on we denote the Hubble rate by $H$ 
and use $H_{inf}$ for the Hubble parameter during inflation.

As the Universe expands, the Hubble parameter 
decreases and eventually gets equal to the ghost mass. 
The corresponding temperature can be readily estimated,
\be
\label{eq:Tosc}
\begin{split}
&H = 1.66 \sqrt{g_*}\frac{T^2_m}{M_P}=m\,,\; \Rightarrow \\
&T_m= \sqrt{\frac{m M_P}{1.66 \sqrt{g_*}}}=3.8\left[\frac{m}{10^{-1}\;\text{eV}}\right]^{1/2}\left[\frac{100}{g_*}\right]^{1/4}\;\text{TeV}\,,
\end{split}
\ee
where $g_*$ denotes the effective number of relativistic degrees of freedom.\footnote{Recall that
for the Standard Model $g_*=106.75$ for $T>200$ GeV.} 

One can show that the curvature perturbation $\R$  
remains conserved outside the horizon after inflation and sources 
the adiabatic mode similar to that of standard cosmology
(see App.~\ref{app:late} for more detail).
As for the non-adiabatic mode of $B$, it remains constant 
at superhorizon scales as long 
as $m< H$.
In the regime $m > H$ this mode starts to oscillate. 
Since the ghost cannot decay, 
the energy of oscillations 
cannot be spent to produce particles and thus 
the energy density of the ghost eventually starts to dominate in the Universe.
Let us roughly estimate when this happens. 
A more rigorous derivation is given in App.~\ref{app:late}.

As we have seen in the previous section, the canonically normalized ghost field 
is given, up to constant, by the Newtonian gravitational potential $\Phi$.
This field is sourced by the both fields $B$ and $\R$ via
\be
\Phi=\R+\HH B\,, 
\ee
see App.~\ref{app:gauge} for more detail. 
Thus, the mode $\Phi$ has two contributions:
the adiabatic mode related to $\R_0$,
and the non-adiabatic mode induced by the 
conserved ghost perturbation $B_0$. Let us focus on 
this non-adiabatic contribution.
At superhorizon scales, before the oscillation started, it is given by
\be
\Phi_{\text{gh}}= \HH B_0\,.
\ee
As one can see, this mode evolves already at superhorizon scales.
The energy density of the ghost is equal to\footnote{Recall that we sacrificed unitarity 
for the positive energy density of the ghost. }
\be
\label{eq:densphi}
\r_{\text{gh}}=m^2 \langle \hat \Phi^2_{\text{gh}}(t,{\bf x})\rangle \,,
\ee
where $\hat \Phi_{\text{gh}}= \sqrt{6}M_P\Phi_{\text{gh}}$ stands for the canonically normalized ghost field. 
Right before oscillations its value is given by
\be
\label{eq:phiin}
\Phi_{\text{gh}}\big|_{m=H} =a(t_m)H(t_m) B_0\,,
\ee
where $t_m$ denotes the transition time at which $m=H(t_m)$.
After the Hubble scale has dropped below the ghost mass, the ghost field 
has the usual solution $\Phi\simeq \text{const}\times \cos(mt+\alpha_0)/a^{3/2}(t)$ (see, e.g.~\cite{Gorbunov:2011zzc}),
where $\alpha_0$ is some phase.
Matching this solution to \eqref{eq:phiin} and using $H(t_m)=m$
we obtain,
\be
\label{eq:phiosc}
\Phi_{\text{gh}}= a(t_m)mB_0\left(\frac{a(t_m)}{a(t)}\right)^{3/2}\cos(mt+\alpha_0)\,.
\ee
Plugging \eqref{eq:phiosc} into Eq.~\eqref{eq:densphi} yields,
\be
\label{eq:rho01}
\begin{split}
\r_{\text{gh}}= & 6M_P^2m^4 a^2(t_m)\langle B_0^2\rangle 
\left(\frac{a(t_m)}{a(t)}\right)^{3}\cos^2(mt+\alpha_0)\,,\\
=& \frac{6M_P^2m^4 a^5(t_m)\cos^2(mt+\alpha_0)}{a(t)^3}\int \frac{dk}{k}\frac{\mathcal{P}_B}{k^2}\,,
\end{split}
\ee
where we used Eq.~\eqref{eq:Bspectr} and the integral above should be performed over coformal momenta. This integral is saturated with the infrared cutoff defined by the Hubble scale,
\be
k_{min}=\HH(\eta)=aH \,.
\ee
Neglecting the tilt of the primordial spectrum $\mathcal{P}_B$,
we obtain the local
energy density of ghosts 
inside the horizon,
\be
\label{eq:rho02}
\begin{split}
\r_{\text{gh}}\simeq & \frac{3M_P^2m^4 a^5(t_m)}{2a(t)^3}
\frac{1}{a(t)^2H(t)^2}\mathcal{P}_B\\
=&\frac{3M_P^2m^4}{2H^2(t)}
\left(\frac{T(t)}{T_m}\right)^5\mathcal{P}_B\,,\\
\end{split}
\ee
where we also averaged over fast oscillations and switched from the scale factor 
to the temperature $T$ in the last equality.

To find the temperature $T_{\text{gh}}$ at which ghosts ``overclose" the Universe 
we equate the energy density \eqref{eq:rho02} 
to the energy density of the background $3H^2M_P^2$,
and use the standard expression for the Hubble rate at radiation domination (see Eq.~\eqref{eq:Tosc}).
This yields
\be
T_{\text{gh}}=\left[\mathcal{P}_B/2\right]^{1/3}(M^*_Pm)^{1/2}
\simeq 260 \left[\frac{H_{inf}}{10^{13}\;\text{GeV}}\right]^{2/3}\left[\frac{m}{10^{-1}\;\text{eV}}\right]^{1/2}\left[\frac{100}{g_*}\right]^{1/4} \;\text{MeV}\,.
\ee
Thus, the ghosts start to dominate in the Universe before primordial nucleosynthesis. 
Assuming that the ghosts are effectively stable,
ghost domination can be avoided only for the masses much below $10^{-1}$ eV,
which are excluded by laboratory constraints discussed in Sec.~\ref{sec:setup}. 
Thus, we conclude that the presence of a stable ghost 
is incompatible 
with viable late-time cosmology.\footnote{
It should be pointed out that the ghost domination
can be avoided in low-scale inflationary scenarios. 
In App. \ref{app:late} 
we show that ghosts never dominate in the Universe if the inflationary Hubble rate $H_{inf}\lesssim 1$ MeV. 
This requires a very unnatural potential for the inflaton (to say nothing of possible issues with reheating),
which is why we do not consider this option here.
}

This result is different from the standard cosmological moduli problem.
For a light free scalar field this problem is much less serious,
and can be avoided for reasonable values of ghost masses and inflationary Hubble rates.
Indeed, let us consider 
a free field $s$ with a mass $m_s$ much smaller than the Hubble 
parameter at inflation.
Fluctuations of this field $\delta s\sim H_{inf}$ are generated at inflation
and are conserved outside the horizon; the corresponding power spectrum is given by 
\be
\mathcal{P}_s=\frac{H_{inf}^2}{(2\pi)^2} \,.
\ee 
In this case the same computation as above would give the following 
current density fraction:
\be
\label{eq:omegascal}
\Omega_{\text{s}} =\frac{\r_{s,0}}{3H_0^2M_P^2}\simeq \frac{1}{3M_P^2}\left[\frac{m_s}{H_0}\right]^{1/2}\Omega_\g^{3/4}\int_{aH} \frac{dk}{k}\mathcal{P}_s \simeq  5\times 10^{-3} 
\left[\frac{H_{inf}}{10^{11}\;\text{GeV}}
\right]^2
\left[\frac{m_s}{0.1\;\text{eV}}\right]^{1/2}\left[\frac{N_e}{60}\right]\,,
\ee
where $H_0$ denotes the current value of the Hubble parameter, $\Omega_\g$ stands for the current 
energy fraction of radiation,  $\r_{s,0}$ stands for the current energy density of the light scalar, and
$N_e$ stands the number of e-foldings (we assumed the spectrum $\mathcal{P}_s$ to be scale invariant).

In our case the situation is quite different. 
The canonically normalized ghost field is not conserved outside the horizon before oscillations.
The conserved perturbation is a
scalar part of the shift vector in the comoving slicing,
which has a different (time-dependent) normalization as compared to a light scalar.

\newpage

\section{Discussion and Conclusions}
\label{sec:concl}

In this paper we studied the creation and evolution of cosmological perturbations in quadratic gravity. 
It has been known for decades that, depending on the quantization prescription,
this theory either suffers from the Ostrogradsky instability or is non-unitary. 
In the first case the theory is pathological both at quantum and classical level,
while in the second case it might still be acceptable from the observational point of view
if non-unitary processes are suppressed.

As it is, non-unitary quadratic gravity cannot be conventionally interpreted
and thus cannot be considered the last word unless the interpretation is found.
However, predictions of the theory will not depend on a particular interpretation 
if the ghost is effectively stable and non-unitary 
processes have negligible rates during the lifetime of the Universe.
We argued that this is the case for ghost masses smaller than 10 MeV.
On the other hand, the lower bound is given by 
laboratory tests of gravity at short scales, which require the ghost to be heavier than 
$0.1$ eV. 
At face value, this yields a rather wide range of ghost masses which is 
phenomenologically viable. 

We showed that inflation generates the usual curvature perturbations 
(which are indistinguishable from that of single-field inflation) along with  
scalar perturbations of the ghost field, 
which are identified with the longitudinal
part of the shift vector
in the comoving slicing gauge.
The vector and tensor perturbations are suppressed.
The scalar ghost fluctuations conserve outside the horizon and 
their energy density 
becomes dominant in the Universe shortly after 
the ghost field started to oscillate.
Thus, our study shows that the presence of a light ghost inevitably leads to a
ghost dominating phase incompatible with the late-time evolution of our Universe. 
This result is independent of any interpretation of non-unitary theories. 

The ghost domination can be avoided 
either by allowing the ghost to decay or by assuming 
that it is not excited during inflation.
In the first case the viability of the model depends crucially on 
the interpretation of processes with negative probabilities.
As for the second option,
in the case of simplest inflationary scenarios 
it implies that the ghost mass must be bigger than $10^{13}$ GeV, 
which violates the bound $m\lesssim 10^{10}$ GeV required for the naturalness of the 
electroweak scale in agravity \cite{Salvio:2014soa}.
Our analysis suggests that one would need a rather exotic inflationary scenario in order 
to have the ghost gapped at inflation and keep the small electroweak scale natural 
in the adimensional gravity model.

Although we have worked in the framework of renormalizable quadratic gravity,
the problem of ghost domination seems to be generic to any theory with the Weyl ghost.
The ghost-dominating phase appears to be quite independent of a particular gravity model, 
and thus
should take place
in any modified gravity containing the Weyl term with the coupling that corresponds to a stable ghost.

\subsection*{Acknowledgments}

We are grateful to D. Levkov, D. Pirtskhalava, 
V. Rubakov, J. Rubio, and S. Sibiryakov for useful discussions and encouragement. 
We are indebted to
F. Bezrukov,  D. Gorbunov, A. Salvio, A. Starobinsky, 
and A. Strumia for valuable comments on the draft.
A. T. acknowledges the support by the Russian Science Foundation grant 14-12-01430 
for the analysis of cosmological perturbations at late times. 
The work of M. I. was supported by the Swiss National Science Foundation. 
M. I. also acknowledges a partial support by the RFBR grant 14-02-00894.

\appendix

\section{Gauge transformations}
\label{app:gauge}

In this section we outline the relations between perturbations in different gauges.
For more detail the reader is advised to consult Refs. \cite{Hu:2004xd,helsinki}.
Let us consider a generic metric of the form,
\be
ds^2=a^2(\tau)\left\{-(1+2\mathcal{A})d\eta^2+2\d_i \mathcal{B} dx^id\eta+\left[(1+2\mathcal{F})\delta_{ij}+2\left(\d_i\d_j-\frac{\delta_{ij}\D}{3}\right)\mathcal{E}\right]dx^i dx^j\right\}\,.
\ee
Then the gauge transformations 
\be
\tilde x^\m =x^\m+\xi^\m 
\ee
generate the following transformations of the metric components and the inflaton, 
\be
\label{eq:gaugetr}
\begin{split}
&\mathcal{\tilde A}= \mathcal{A}-(\xi^0)'-\HH \xi^0\,,\\
&\mathcal{\tilde F}= \mathcal{F}+\frac{1}{3}\Delta \xi-\HH \xi^0\,,\\
&\mathcal{\tilde B}= \mathcal{ B}+\xi'+\xi^0\,,\\
& \mathcal{\tilde E}=\mathcal{E}+\xi\,,\\
& \tilde \varphi=\varphi-\phi'\xi^0\,,
\end{split}
\ee
where $\xi\equiv -\d_i \xi^i/\D$.
For a generic matter field the perturbations of its density and velocity potential transform as
\be 
\begin{split}
&\delta \tilde \rho_{tot}=\delta\rho_{tot}-\r'_{tot}\xi^0\,,\\
&[(p+\r)\tilde v]_{tot}=[(p+\r) v]_{tot}+(p+\r)_{tot}\xi'\,.
\end{split}
\ee

The relations between perturbations in the conformal 
Newtonian ($\varphi, \Phi,\Psi$, see Eq.~\eqref{metric}) and 
the uniform inflaton ($A,B,\R$, see Eq.~\eqref{eq:comg}) gauges are given by
\be
\label{eq:eq:comtoNew}
\begin{split}
&\Phi= \R+\HH B\,,\\
&\Psi= A+B'+\HH B \,,\\
& \varphi=\phi' B\,.
\end{split}
\ee

Using the transformation rules \eqref{eq:gaugetr}
one can define the perturbation $B$ in a gauge invariant way,
\be
\begin{split}
 B^{(\text{g.i.})}=
 \begin{cases}
\mathcal{B}-\mathcal{E}'+{\delta \r_{tot}}/{\r'_{tot}} ~~~~\text{in general}\,,\\
\mathcal{B}-\mathcal{E}'+{\varphi}/{\phi'}~~~~~~~~~~\text{at inflation}\,.
 \end{cases}
\end{split}
\ee

\section{Inflationary perturbations in the uniform inflaton gauge}
\label{app:comoving}

In the main text we have shown that the conformal Newtonian gauge is 
unstable during inflation since the perturbations undergo an uncontrolled 
growth outside the horizon.  We argued that perturbations behave well in the 
uniform inflaton gauge, and extracted the corresponding mode functions from 
those obtained in the Newtonian gauge by means of the gauge transformation 
\eqref{eq:eq:comtoNew}.
We argued that this procedure is consistent even though the Newtonian gauge is unstable.
In this section we verify this point by a direct calculation in the uniform inflaton gauge.

In the uniform inflaton (comoving slicing) gauge the scalar perturbations take the form
\begin{equation}
\label{eq:comg}
\phi=\phi(\eta)\,,\quad g_{00}=-a^2(1+2A),~ g_{0i}= a^2 \partial_i B,~g_{ij}=a^2\delta_{ij}(1+2 \R)
\,.
\end{equation}
The total quadratic action \eqref{eq:action} then reads,
\be
\begin{split}
S^{(2)}=&M_P^2\int d^3xd\eta a^2\Bigg[-\R\Delta\R -3{\R'}^2 +6\HH{\R'}A
-2A\Delta\R-2\Delta B(\HH A-\R')\\
&-\left(\HH'+2\HH^2\right)A^2 -\frac{1}{3a^2m^2} \Delta^2(A-\R +B')^2
\Bigg]\,.
\end{split}
\ee
The variation of this action with respect to $A$ produces a constraint
\be
\label{eq:constCOM}
6\HH \R'
 -2\Delta \R 
 -2\Delta B\HH
 -2A(\HH'+2\HH^2)
 -2\frac{\Delta^2}{3a^2m^2}(A-\R+B')=0\,.
\ee
As in the Newtonian gauge, we observe that 
the above equation can be solved in two different limits: 
the ``Weyl term domination", characterized by inequality 
\[
k_{gh}\equiv k^2/(a(\eta)m)\gg k,\HH\,,
\]
and the ``Einstein term domination", at which the above inequality changes to 
the opposite one.
Solving the constraint 
\eqref{eq:constCOM}
in the regime of the Weyl term domination ($k_{gh}\gg k,\HH$)
and plugging the solution back yields the following
quadratic action:
\be
\label{eq:act}
\begin{split}
S^{(2)}\Big|_{k_{gh}\gg \HH,k}=& M_P^2\int d^3x d\eta \; a^2\big[ -3\R'^2 -3\R \Delta\R -4(\HH'+2\HH^2) \R^2 -(\HH'+2\HH^2)(B'^2+B\Delta B)\\
& -6\HH \R' B' -6\HH \R \Delta B+2(\HH'+2\HH^2)\R B'\big]\,.
\end{split} 
\ee
In order to quantize the fields $\R$ and $B$ we have to diagonalize the above action
inside the Hubble horizon. This amounts to introducing new fields $\tilde \chi$ and $\tilde \Phi$ via
\begin{subequations}
\label{eq:gaugeback}
\begin{align}
& \tilde \chi= \phi' B \,,\\
 &\tilde \Phi= \sqrt{6}M_P(\R+\HH B)\,,
\end{align}
\end{subequations}
for which the action \eqref{eq:act} takes the following form inside the horizon at the leading order in the slow roll parameters,
\be
\label{eq:act-diag}
\begin{split}
S^{(2)}\Big|_{k_{gh}\gg k\gg \HH}=& \frac{1}{2}\int d^3x d\eta \; a^2\big[
(\tilde \chi'^2 -(\d_i \tilde \chi)^2) -(\tilde \Phi'^2 -(\d_i\tilde \Phi)^2)\big]\,,
\end{split} 
\ee
Looking at Eq.~\eqref{eq:eq:comtoNew} we see that 
Eq.~\eqref{eq:gaugeback} is nothing but a gauge transformation, which brings us 
back to the Newtonian gauge; the field $\tilde \chi$ coincides with the inflaton, 
the field $\tilde \Phi$ - with the canonically normalized Newtonian potential.
These fields represent two independent (inside the horizon) degrees of freedom.
This implies that quantization can conventionally be  
performed only in the Newtonian gauge.

Now we check that all perturbations in the uniform inflaton gauge get frozen outside the horizon.
The action \eqref{eq:act} implies the following equations of motion,
\be
\label{eq:eomADM}
\begin{split}
&\R''+2\HH\R'-\Delta \R+\frac{4}{3}(\HH'+2\HH^2)(B'-\R)+\HH (B''-\Delta B)=0\,, \\
& (\HH'+2\HH^2)(B''-\Delta B) +(\HH''+6\HH'\HH+4\HH^3) (B' -\R) +2(\HH'+2\HH^2) \R'
+3\HH (\R'' -\Delta \R)=0\,.
\end{split}
\ee

Naively the above equations are regular in the limit of exact
de Sitter space. In this case the two field equations are equivalent, 
which is a consequence of the fact that $\R$
becomes a gauge mode, 
and thus we are left only with only one ghost degree of freedom. 

Now let us find a non-trivial solution to the equations Eqs.~\eqref{eq:eomADM} in the slow-roll limit.
Upon summing and subtracting these equations from each other they can be brought to the following form,
\be 
\begin{split}
&(\HH'-\HH^2)\Big[B''-\Delta B +2\R'+(B'-\R) \frac{\HH''+2\HH\HH'-4\HH^3}{\HH'-\HH^2}\Big]=0\,,\\
&\frac{(\HH'-\HH^2)}{\HH'+2\HH^2}\Big[\R''-\Delta \R-(B'-\R)\frac{3\HH \HH''-4\HH'^2+2\HH^2\HH'-4\HH^4}{3(\HH'-\HH^2)}\Big]=0\,.
\end{split}
\ee
Note that no slow-roll approximation has been made so far. 
Now assume that the modes are outside the Hubble horizon, 
hence the terms with spacial laplacians can be 
neglected.
Next we combine the above equations in order to obtain a single homogeneous equation on $\R$,
and assume a quasi-de Sitter background with 
$a\propto \eta^{1/(\epsilon_1-1)}$. One obtains,
\be
\begin{split}
 & \R''' +\R''\frac{4}{(\epsilon_1-1)\eta}+\frac{2(3-\epsilon_1)}{(\epsilon_1-1)^2\eta^2}\R'=0\,.
 \end{split}
\ee
The solution at the leading order in $\epsilon_1$ 
is given by a constant mode and two modes that vanish in the limit $\eta \to 0$,
\be 
\label{eq:Rmodes}
\R=\left\{ \text{const},\quad \eta^{4+2\epsilon_1},\quad \eta^{3+2\epsilon_1} \right\}\,.
\ee

The constant mode $\R=\R_0$ implies the following solution for the mode $B$,
\be
\label{eq:supz}
\begin{split}
&\R=B'=\R_0=\mathrm{const}\,, \\
&\Rightarrow \quad  B=\R_0 \eta + B_0\,,
\end{split}
\ee
where $B_0$ is a constant. 
The decaying modes of $\R$ \eqref{eq:Rmodes} 
source only decaying solutions for $B$.

In the limit of the Einstein term domination 
($k_{gh}\ll \HH$) one
finds the following action at first order in $ k^2/\HH^2,\; k^2_{gh}/\HH^2,\;\epsilon_1$ (see Eq.~\eqref{eq:eps}),
\be
\begin{split}
S^{(2)}\Big|_{k_{gh}\ll \HH}=& M_P^2\int d^3x d\eta \; a^2\bigg[ 
\epsilon_1 \R'^2
-\frac{\Delta^2}{3m^2a^2}\left(B'+\frac{3\HH \R'}{\HH'+2\HH^2}-\R\right)^2
\bigg]\,.
\end{split} 
\ee
It is trivial to show that the solution \eqref{eq:supz} holds in
this case too.

To sum up, we have shown that the solution \eqref{eq:supz} coincides with the 
one obtained by gauge transforming 
the solutions from the Newtonian gauge (see Eqs.~\eqref{eq:zeta1} and \eqref{eq:B1}).

\section{Cosmological perturbations at late times}
\label{app:late}

\subsection{Main equations: qualitative analysis}

In this section we consider the dynamics of linear cosmological perturbations after inflation and reheating.
Previous results suggest us to work from the beginning in the comoving 
slicing gauge where we expect perturbations to be free of unphysical instabilities.
For simplicity, we will focus on the case of a single component matter, 
which is known to be a very good approximation since it describes 
perturbations of a component that dominates at a given cosmological epoch.  
This approximation will be enough for our purposes.
In practice, we use the energy-momentum tensor  
\be
T_{\mu\nu} =(\rho+p)v_{\mu}v_{\nu}+pg_{\mu \nu}\,,
\ee
assuming a constant equation of state $p=c^2_s\rho$.
Since the Weyl term vanishes on the FRW background,
assuming the metric \eqref{eq:backgr} and the ansatz 
$\rho=\bar \rho (\eta)$, $p=\bar p (\eta)$ 
we obtain the usual set of the Friedman and continuity equations,
\be 
\begin{split}
& 3M_P^2 \HH^2=a^2 \bar \rho\,, \\
& \bar \rho'+3\HH (1+c_s^2)\bar \rho=0\,.
\end{split}
\ee

At the level of perturbations,
we assume the metric on the comiving slicing \eqref{eq:comg} 
and the following perturbations of the energy-momentum tensor,
\be
\label{eq:pertEMT}
\begin{split}
& \rho=\bar \rho+\delta \rho \,,\\
& p=\bar p +\delta p \,,\\
& v_\mu=(-a(1+\Psi),0)\,,\\
& v^\mu=\frac{1}{a}((1-\Psi),-\d_iB)\,,\\
\end{split} 
\ee
where we have used the definition of the comoving slicing gauge
$\delta T_i^0=0$ \cite{Hu:2004xd}.
Substituting the metric \eqref{eq:comg0} 
and the energy momentum tensor components \eqref{eq:pertEMT}
into \eqref{eq:eqgen} we find the following set of modified
Einstein equations,
\begin{subequations}
\label{eq:einst}
\begin{align}
& -\Delta (\R +\HH B)+3\HH \R' 
-3\HH^2 A-\frac{\Delta^2}{3a^2m^2}(A+B'-\R)=\frac{a^2}{2M_P}\delta \r\,,\\
&-\R'+\HH A -\frac{\Delta}{3a^2m^2}(A'+B''-\R')=0\,,\\
\nonumber
& -\R''-2\HH \R' +\HH A'
+\frac{1}{3}\Delta (\R+B'+A+2\HH B)\\
& ~~~~~~~~~~~~~~~~~~~~~~~~+(\HH^2+2\HH')A
-\frac{\Delta^2}{9m^2a^2}(A+B'-\R)
-\frac{a^2}{2M_P^2}\delta p =0\,,\\
& A+B'+2\HH B+\R+\frac{1}{a^2m^2}(\d_\eta^2-\frac{\Delta}{3})(A+B'-\R)=0\,.\end{align} 
\end{subequations}
The conservation of the energy momentum tensor yields two more equations,
\begin{subequations}
\label{eq:cont}
\begin{align}
&A=-\frac{\delta p}{(\bar p+\bar \r)}\,,\\
&A'=c_s^2\Delta B+3c_s^2 \R'\,,
\end{align} 
\end{subequations}
where in order to obtain the last equation we used $\delta p=c_s^2\delta \rho$.

Upon inspection of Eqs.~\eqref{eq:einst} and \eqref{eq:cont}
one notices that there are three equations that form a closed system
for the variables $B, A$ and $\R$,

\begin{subequations}
\label{eq:mastsys}
\begin{align}
& ~~~ \R'-\HH A=-\frac{\Delta}{3 a^2 m^2}(B''+A'-\R')\,,\\
& ~~~ A+B'+2\HH B+\R+\frac{1}{a^2m^2}\left(\d_\eta^2-\frac{\Delta}{3}\right)(A+B'-\R)=0\,,\\
&
\begin{cases}
A'=c_s^2(\Delta B-3 \R') ~~~~\mathrm{if}~~~ c_s^2\neq 0\,,\\
 A=0~~~~~~~~~~~~~~~~~~~~~~\mathrm{if}~~~ c_s^2= 0\,.
 \end{cases}
\end{align}
\end{subequations}

Now let us study Eqs.~\eqref{eq:mastsys} 
in the superhorizon limit where we can neglect spatial laplacians.
We also assume the scale factor $a(\eta)\propto \eta^p$.
This yields in the following system:
\begin{subequations}
\label{eq:mastsyssuper}
\begin{align}
\label{eq:sup1}
& ~~~ \R'-\frac{p}{\eta} A=0\,,\\
\label{eq:sup2}
& ~~~ A+B'+2\frac{p}{\eta}B+\R+\frac{1}{a^2m^2}(A''+B'''-\R'')=0\,,\\
&
\begin{cases}
A'=-3c_s^2\R' ~~~~\mathrm{if}~~~ c_s^2\neq 0\,,\\
 A=0~~~~~~~~~~~~~\mathrm{if}~~~ c_s^2= 0\,.
 \end{cases}
\end{align}
\end{subequations}
The mode $\R=\R_0=$ const induces the standard adiabatic mode,
\be
\label{eq:superhin}
\begin{split}
&A=0\,,\\
&B=-\frac{\R_0 \eta}{1+2p}\,.
\end{split} 
\ee
Remarkably, this solution is valid regardless the value of the 
ghost mass, i.e. for both regimes $H\gg m$ and $H\ll m$.

In order to study the dynamics of the isocurvature mode of $B$
we consider Eq.~\eqref{eq:sup2} with a vanishing source 
$\R=0$ (which also implies $A=0$).
We also assume, for definiteness, the radiation domination stage with $p=1$. In this case 
Eq.~\eqref{eq:sup2} reduces to 
\be 
\label{eq:Bmaster}
B'''+a^2m^2\left(B'+\frac{2}{\eta}B\right)=0\,.
\ee
In the limit $m\to 0$ the above equations has three solutions, 
$B=$ const, $B\propto \eta$ and $B\propto \eta^2$,
but the last two solutions are negligible in the limit $\eta \to 0$.
This implies that 
the inflationary mode $B=B_0$ is conserved.

Once the Hubble rate drops below the ghost mass, the field $B$ starts to oscillate. 
Eq.~\eqref{eq:Bmaster} can be rewritten as
\be 
\label{eq:na}
\left[\Phi_{\text{gh}}''+\frac{2}{\eta} \Phi_{\text{gh}}'+a^2m^2\Phi_{\text{gh}}\right]'=0
\,,\quad \text{where}
\quad \Phi_{\text{gh}}=\HH B=\frac{B}{\eta}\,,
\ee
so it takes the form of a usual Klein-Gordon-Fock equation 
for the massive field at superhorizon scales. 
Eq.~\eqref{eq:na} describes the behavior of 
a contribution to the Newtonian potential that is 
induced by a non-adiabatic mode of $B$.
The initial conditions for an oscillating solution of Eq.~\eqref{eq:na} 
are defined by a constant $B_0$.  

\subsection{Ghost energy density}

As we have seen in the previous section, the field $B$ has a constant non-adiabatic 
mode at superhorizon scales as long as $m<H$. Once the Hubble rate gets smaller than
the ghost mass, the field $B$ starts to oscillate. Since the interaction of the ghost with 
matter is suppressed, the production of particles does not occur and 
the oscillation energy of the ghost field starts to dominate the cosmic expansion. 
Neglecting the coupling to matter and using Eqs.~\eqref{eq:weyl2} and \eqref{eq:gr2}
it can be shown that at the leading order in the Hubble rate the quadratic action 
of the ghost takes the following form,\footnote{
Formally, in deriving 
the quadratic action~\eqref{eq:gr2} we made use 
of the background equations of motion for the inflaton. 
In the case of an arbitrary matter content the result will be the same, which can easily be shown
e.g. by writing the macroscopic matter action 
along the lines of Ref.~\cite{mukhanov}.
}
\be
\label{eq:actP}
S=-3M_P^2\int d^3x dt\; a^3\left(\dot \Phi_{\text{gh}}^2-\frac{(\d_i\Phi_{\text{gh}})^2}{a^2}-m^2\Phi^2_{\text{gh}}\right)\,.
\ee
The non-adiabatic constant mode of $B$ defines the following asymptotics of $\Phi_{\text{gh}}$
in the limit $t\to 1/m$:
\be
\label{eq:incond}
\Phi_{\text{gh}}=\HH B_0\to \frac{a_{\gamma} }{2t^{1/2}}B_0\,,
\ee
where we used the scale factor $a(t)=a_\gamma \cdot t^{1/2}$ at radiation domination.
A concrete expression for $a_\gamma$ will not be relevant for us.
The action \eqref{eq:actP} yields the following equation of motion:
\be
\label{eq:phieq}
\ddot \Phi_{\text{gh}} +3 H \dot \Phi_{\text{gh}}+\left(-\frac{\Delta}{a^2}+m^2\right)\Phi_{\text{gh}}=0\,.
\ee
For cosmologically relevant modes the mass term always dominates, 
hence one can safely neglect the spatial laplacian. Then the solution of 
Eq. \eqref{eq:phieq} satisfying the initial behavior \eqref{eq:incond}
is given by, 
\be
\label{eq:Bessel}
\Phi_{\text{gh}}= -\frac{\pi  a_\gamma B_0 \sqrt[4]{m}}{2 \sqrt[4]{2} \Gamma \left(\frac{1}{4}\right)}\frac{Y_{1/4}(mt)}{t^{1/4}}\,,
\ee
where $Y_{1/4}(x)$ is the Bessel function of the second kind, and $\Gamma(x)$ is the Euler gamma function.
At $t\gg 1/m$ the solution \eqref{eq:Bessel} has the following asymptotic:
\be
\label{eq:amplitude}
\Phi_{\text{gh}}\approx \frac{\pi  a_\gamma B_0 \sqrt[4]{m}}{2 \sqrt[4]{2} \Gamma \left(\frac{1}{4}\right)}\frac{1}{t^{1/4}}\sqrt{\frac{2}{\pi(mt)}}\cos\left(mt+\frac{\pi}{8}\right)=
\frac{\sqrt{\pi}}{2^{3/4}\Gamma \left(\frac{1}{4}\right)}B_0\frac{a_\g^{5/2}}{m^{1/4}a(t)^{3/2}}\cos\left(mt+\frac{\pi}{8}\right)\,.
\ee
In order to estimate the energy density of the ghost we rewrite the action \eqref{eq:actP} in comoving 
coordinates $\tilde x\equiv a(t)x$, 
\be
\label{eq:actP2}
S=-3M_P^2\int d^3\tilde x dt\;\left(\dot \Phi^2_{\text{gh}}-{(\tilde \d_i\Phi_{\text{gh}})^2}-m^2\Phi^2_{\text{gh}}\right)\,,
\ee
and compute the corresponding energy functional using the decomposition \eqref{eq:dec}
and the negative-norm prescription \eqref{eq:negative}, which yields,
 \be 
 \label{eq:rhogw}
E=\int d^3\tilde x \rho_{\text{gh}}=\int d^3\tilde x\left[3M_P^2\int \frac{dq}{q}\Phi^2_{\text{gh}}(q)\;\left(\omega^2_q+q^2+m^2\right)\right]\,,
 \ee
 where $\omega_q$ and $q$ are the comoving frequency and momentum. 
 Plugging the amplitude \eqref{eq:amplitude} in Eq.~\eqref{eq:rhogw} 
 and using the dispersion relation $\omega^2_q=m^2+q^2\approx m^2$ one finds,
 \be 
  \begin{split}
 \label{eq:rhogw2}
 \rho_{\text{gh}}=&6m^2 M_P^2\int \frac{dq}{q}\Phi^2_{\text{gh}}(q)\,,\\
 =& \frac{\pi}{2^{3/2}\Gamma^2 \left(\frac{1}{4}\right)}\frac{a_\g^{5}}{m^{1/2}a(t)^{3}}6m^2 M_P^2\cos^2\left(mt+\frac{\pi}{8}\right) \times \left[\int \frac{dq}{q}B_0^2(q)\right]\,,
 \end{split}
 \ee 
By introducing the transition time 
 \[
 t_m:\,H(t_m)=m\,,
 \]
Eq.~\eqref{eq:rhogw2} can be rewritten as
  \be 
  \begin{split}
 \label{eq:rhogw4}
 \rho_{\text{gh}}
 = \frac{\pi}{\Gamma^2 \left(\frac{1}{4}\right)}\frac{6 m^4 M_P^2a^{5}(t_m)}{a(t)^{3}} \cos^2\left(mt+\frac{\pi}{8}\right)\times \left[\int \frac{dq}{q}B_0^2(q)\right]\,.
 \end{split}
 \ee
This expression coincides with our rough estimate \eqref{eq:rho01} up to a factor $\pi/\Gamma^2 \left(\frac{1}{4}\right)\approx 0.24$.

In Sec.~\ref{sec:late} we showed that for reasonable values of ghost masses $m\gtrsim 0.1$ eV and 
Hubble rates at inflation $H\sim 10^{13}-10^{16}$ GeV
the ghost energy density becomes dominant before 
primordial nucleosynthesis. 
Let us assume now that the inflation happened at very low energies and
estimate for which value of the inflationary Hubble rate the 
ghost domination does not happen.
First relevant observation is that the solution \eqref{eq:amplitude}
holds at matter domination, which implies that the expression \eqref{eq:rhogw4}
does not change either.
Then, using the expression for the scale factor at radiation
domination $a=a_\g t^{1/2}=\sqrt[4]{\Omega_\g}\sqrt{2 H_0 t}$ 
we find the current energy fraction of the ghosts,
\be 
  \begin{split}
 \label{eq:rhogw0}
\Omega_{\text{gh}}\equiv \frac{\rho_{\text{gh},0}}{3M_P^2H_0^2}
 =& \frac{\pi}{\Gamma^2 \left(\frac{1}{4}\right)}\mathcal{P}_B\Omega_\g^{5/4}
 \left[\frac{m}{H_0}\right]^{3/2}\\
 \simeq & 0.1\left[\frac{H_{inf}}{4\;\text{MeV}}\right]^2\left[\frac{m}{10^{-1}\;\text{eV}}\right]^{3/2}\,,
 \end{split}
 \ee
 where $H_{inf}$ denotes the Hubble parameter at inflation, 
$\Omega_\g$ - the current energy fraction of radiation,
$H_0$ - current Hubble rate.
This expression should be contrasted with Eq.~\eqref{eq:omegascal}.
Compared to a light moduli,
given the same masses, the ghost field requires a significantly lower scale of inflation 
in order not to ``overclose" the Universe.


\end{document}